\documentclass[acmsmall,screen]{acmart}

\setcopyright{rightsretained}
\copyrightyear{2024}
\acmYear{2024}
\acmDOI{XXXXXXX.XXXXXXX}

\usepackage{graphicx} 
\usepackage{subcaption}
\usepackage{booktabs}
\usepackage{hyperref}
\usepackage{listings}
\usepackage{float}
\usepackage{tabularx}
\usepackage{wrapfig}
\usepackage{tikz-imagelabels}
\usepackage{natbib}
\setcitestyle{square,citesep={,}}

\hypersetup{hidelinks,linkcolor = black}

\begin{document}

\title{Inferring Alt-text For UI Icons With Large Language Models During App Development}

\author{Sabrina Haque}
\affiliation{%
  \department{Department of Computer Science and Engineering}
  \institution{University of Texas at Arlington}
  \city{Arlington}
  \state{Texas}
  \country{USA}}
\email{sxh3912@mavs.uta.edu}

\author{Christoph Csallner}
\affiliation{%
  \department{Department of Computer Science and Engineering}
  \institution{University of Texas at Arlington}
  \city{Arlington}
  \state{Texas}
  \country{USA}}
\email{csallner@uta.edu}

\lstset{
    basicstyle=\ttfamily\scriptsize,
    breaklines=true,
    frame=single,
    numbers=left,
    numberstyle=\tiny,
    keywordstyle=\color{blue},
    stringstyle=\color{red},
    commentstyle=\color{green!60!black},
    showstringspaces=false,
    captionpos=b
}

\newcommand{\code}[1]{\lstinline[basicstyle=\ttfamily\small]{#1}}

\newcommand{\toolName}{\textsc{IconDesc}} 

\begin{abstract}

Ensuring accessibility in mobile applications remains a significant challenge, particularly for visually impaired users who rely on screen readers. User interface icons are essential for navigation and interaction and often lack meaningful alt-text, creating barriers to effective use. Traditional deep learning approaches for generating alt-text require extensive datasets and struggle with the diversity and imbalance of icon types. More recent Vision Language Models~(VLMs) require complete UI screens, which can be impractical during the iterative phases of app development. To address these issues, we introduce a novel method using Large Language Models (LLMs) to autonomously generate informative alt-text for mobile UI icons with partial UI data. By incorporating icon context, that include class, resource ID, bounds, OCR-detected text, and contextual information from parent and sibling nodes, we fine-tune an off-the-shelf LLM on a small dataset of approximately 1.4k icons, yielding \toolName{}. In an empirical evaluation and a user study \toolName{} demonstrates significant improvements in generating relevant alt-text. This ability makes \toolName{} an invaluable tool for developers, aiding in the rapid iteration and enhancement of UI accessibility.
\end{abstract}

\ccsdesc[500]{Software and its engineering~Software usability}
\ccsdesc[500]{Human-centered computing~Accessibility systems and tools}
\ccsdesc[300]{Human-centered computing~Empirical studies in accessibility}

\keywords{Mobile app development, developer support, visual impairment, fine-tuning, icon context, icon alt-text}

\maketitle

\section{Introduction}

Many mobile apps assume that their users will interact with the app based on the visual appearance of the app's small interactive graphical UI elements, i.e., its icons and image buttons. To support alternative (non-visual) user-app interaction styles, the major mobile platforms also allow each app to describe the meaning of its icons via text (i.e., accessibility metadata such as alt-text). But many apps do not provide meaningful descriptions for all of their relevant GUI elements~\cite{ross2018examining,alshayban2020accessibility}. Recent work has made great progress toward automatically inferring such textual descriptions via deep learning and large language models. But many challenges remain, including supporting the long tail of rare GUI elements and supporting engineers during development where full screen information may not yet be available.

Today billions of people rely on mobile apps for a wide variety of tasks. Governments have enacted regulations such as the Americans with Disabilities Act (ADA) and Section 508 of the U.S. Rehabilitation Act~\cite{ada,508_act}. These regulations mandate accessibility standards. The Screen Reader User Survey conducted by WebAIM in 2017 reported that 90.9\% of blind or visually impaired respondents relied on screen readers on smartphones~\cite{screenreader_survey}. Yet research highlights persistent accessibility issues in mobile apps, creating barriers that impede their efficient use by people with various injuries and disabilities~\cite{ross2017epidemiology,ross2018examining,yan2019current,alshayban2020accessibility}. Central to these challenges are the graphical user interface (GUI) elements lacking meaningful alt-texts.

Inferring the meaning of an arbitrary icon remains a fundamentally hard problem. We can only make a limited number of observations yet have to summarize and generalize these observations to a description that captures the icon's meaning in all possible app usage scenarios. To make this generalization, prior work has leveraged a variety of types of observation (including screenshots, view hierarchy information, and app runtime behavior). We are not aware of work however that supports an engineer during development with high-quality alt-texts when the full screen information is not available yet. This additional lack of information makes this task even harder.

At a high level, recent work falls into one of the following two broad categories. In the first category are traditional deep-learning techniques that are specialized for inferring GUI element alt-texts~\cite{chen2020unblind,mehralian2021data}. While they paved the way for subsequent work, these pioneering approaches are trained on relatively small data sets.
In the second category are modern models including large multi-modal models and large language models~\cite{gpt4,chatgpt,Gemini-1.5}. While these models have been trained on much larger datasets, they either are not fine-tuned on domain-specific UI tasks~\cite{jiang2023iluvui} or do not perform well on partial screens.

To address the challenges previous approaches face, this paper introduces a novel application of off-the-shelf large language models to generate context-aware alt-texts---without requiring complete screen information. This approach is especially beneficial during the early phases of app development, where screen data (such as screenshots, UI elements, view hierarchies, and UI code) are often incomplete. This paper implements this approach in the \toolName{} tools and compares \toolName{} with state-of-the-art (SOTA) approaches.

Besides improving on SOTA approaches in terms of commonly-used automated metrics and in a user study, \toolName{} also only requires a small dataset (for fine-tuning), which compares favourably to the (at least an order of magnitude) larger datasets that are needed to train specialized deep-learning approaches.

To summarize, the paper makes the following contributions.
\begin{itemize}
    \item The paper introduces a novel pipeline technique for generating alt-text for UI icons from partial screen data. The pipeline infers an icon-only description via a large multi-modal model and fine-tunes a large language model to generate alt-texts.
    \item The paper implements the technique in \toolName{} and compares \toolName{} empirically with specialized deep-learning approaches and modern SOTA models, achieving top scores in both automated metrics and a user study.
    \item All code, data, and configurations we used (including for fine-tuning) are available~\cite{our_code}.
\end{itemize}

\section{Background}

Mobile apps consist of various screens or activities filled with user interface (UI) elements designed to facilitate user interaction. Among these elements, interactive components such as buttons, checkboxes, clickable images, and icons provide concrete engagement points for users to initiate actions and navigate app features. Icons' widespread use in mobile apps can be attributed to their ability to convey information effectively while consuming minimal screen space.

\subsection{Accessibility Challenges: Inferring the Meaning of Small UI Elements}

Despite their utility, a significant challenge with these small interactive UI elements is that they often lack a meaningful alt-text label~\cite{fok2022large,ross2018examining}. Meaningful alt-text labels are crucial for accessibility technologies such as screen readers~\cite{talkback, voiceover}, so not having meaningful alt-texts impedes the navigation and interaction for users relying on assistive technologies. For example, research has highlighted the pervasive issues of missing alt-text across Android apps, underscoring the need for automated generation of descriptive alt-texts~\cite{fok2022large, ross2018examining}.

\begin{wrapfigure}{r}{.5\linewidth}
\centering
\includegraphics[width=.48\linewidth]{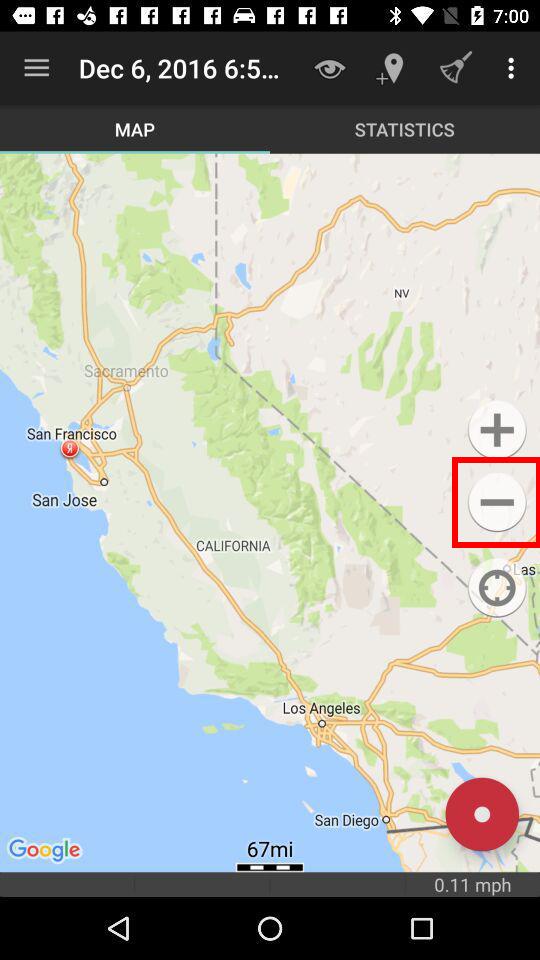}
\includegraphics[width=.48\linewidth]{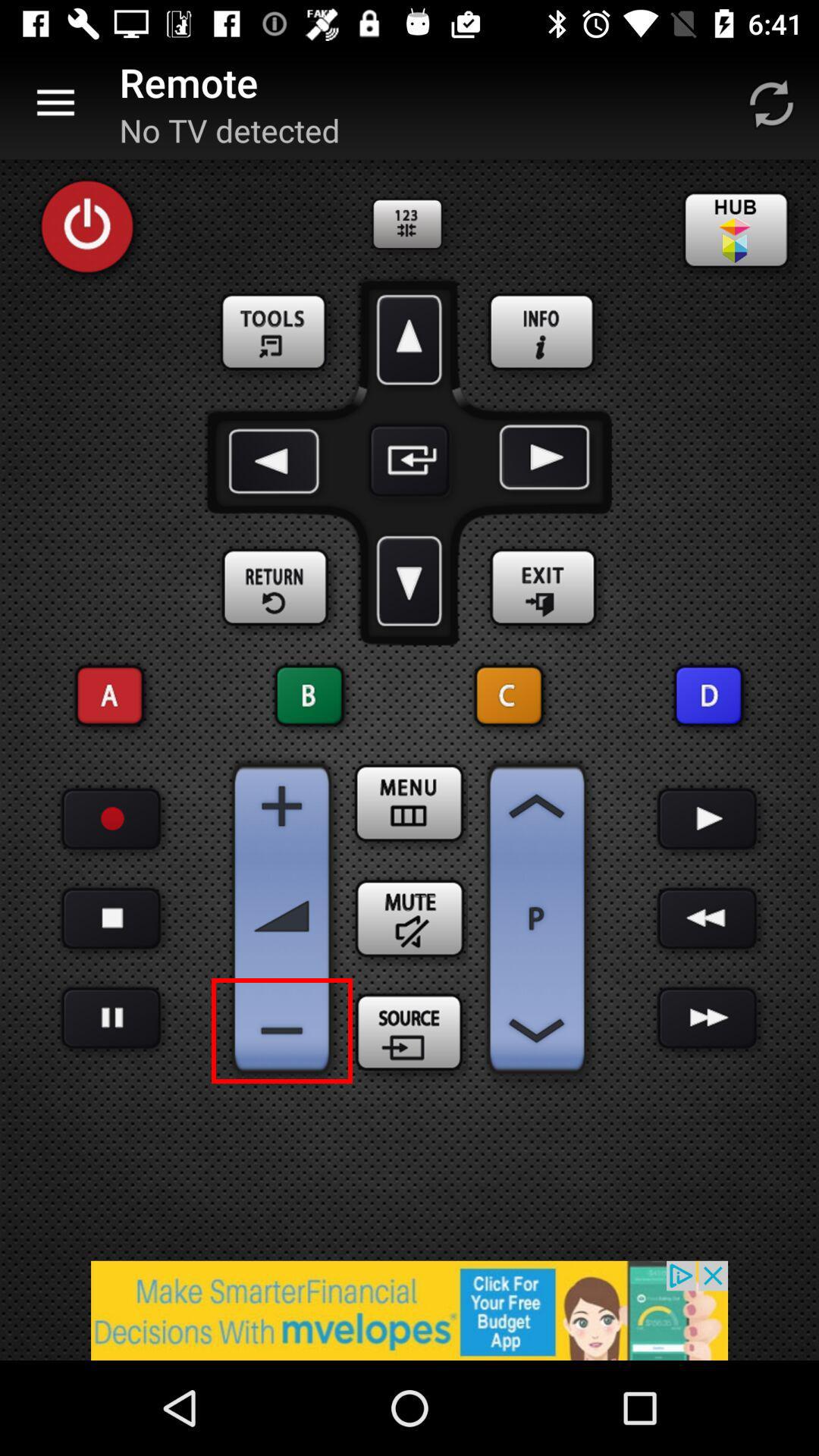}
\caption{Example zoom out (left) vs. lower volume (right) minus buttons in Rico.}
\label{fig:icon_context}
\end{wrapfigure}

For inferring alt-text, a sample challenge is that a given kind of icon may represent different functionalities depending on the context. As an example from the Rico dataset~\cite{deka2017rico}, Figure~\ref{fig:icon_context} shows two app screens. The left screen uses a ``minus'' icon for zooming out while the right screen uses a similar ``minus'' icon for decreasing the volume. To support such cases we thus need an alt-text inference tool to interpret an icon's pixels in the icon's screen context.

\subsection{Large Language and Vision Models}

Large language models (LLMs) have become powerful tools in natural language processing, demonstrating exceptional performance on a variety of text-based tasks. These transformer-based~\cite{vaswani2017attention} models, equipped with billions of parameters and trained on extensive datasets, can comprehend and generate human language proficiently~\cite{chowdhery2023palm,chatgpt}.

Beyond processing text, vision-language models (VLMs) are trained on both text and images, letting them address complex tasks such as captioning and scene-text understanding~\cite{llava,li2021vut,chen2022pali}. Recent more general models support multiple modalities (text, images, etc.) and are sometimes also referred to as multi-modal models~\cite{gpt4,Gemini-1.5}.

The key attribute of all of these transformer-based models is the amount (and diversity) of data used in their training. So we distinguish large and state-of-the-art models (the latter now require large amounts of training data) from the broader group of models of all sizes.

In any case, a model is trained on diverse data sets, which equips a model to handle a wide variety of tasks across domains.

There are several options for adapting such a model to new tasks. 
First, zero-shot learning allows the model to perform a task it has never seen before purely based on the instructions given in the prompt, which can be difficult for unseen tasks. Specifically, prompting involves using natural language instructions within the model input, to guide the LLM toward the desired output. These prompts can range from simple question formulations to elaborate demonstrations and background information. 
Building on prompting, in-context learning (aka few-shot learning) incorporates a few examples of the desired task within the prompt~\cite{brown2020language}. This provides the LLM with concrete demonstrations of the desired behavior, enhancing its task-specific understanding.

While prompting and in-context learning are powerful methods for adapting models to new tasks, there are cases where these approaches might not suffice. Fine-tuning is a traditional approach to adapting a model for specific tasks via additional training on a small task-specific dataset, thereby updating the model's many parameters. Fine-tuning can be very costly and is thus sometimes not worth the effort. Platforms such as OpenAI provide APIs for fine-tuning on custom datasets~\cite{openai_finetuning}.

\subsection{Alt-text Generation: LabelDroid, Coala, Widget Captioning, etc.}

LabelDroid~\cite{chen2020unblind} marked the first attempt based on deep learning for automating alt-text generation for UI icons. LabelDroid utilizes only icon images to generate alt-text in the style of image captioning. While groundbreaking, LabelDroid did not take into account context information. The subsequent Coala deep-learning model~\cite{mehralian2021data} added both visual and contextual information from the app’s view hierarchy during training. This approach resulted in improvements in the relevance and accuracy of generated alt-texts.

In a similar deep-learning approach, Li et al. introduced Widget Captioning~\cite{li2020widget}, which not only focuses on icons but broadens the scope to include other UI elements (buttons, check boxes, image views, etc.). By utilizing both visual and structural data of UI components, this method efficiently generates descriptive captions. This broader approach was supported by a dataset of human-annotated widget captions, which serves as a foundation for our research.

Instead of applying traditional deep learning, recent work adapts VLMs to this task. Models such as Pix2Struct and PaliGemma have shown notable success in widget captioning tasks, inferring descriptive text for UI elements based solely on visual input.
For example, Pix2Struct leverages the rich visual and textual data from web pages by parsing masked screenshots into simplified HTML, which facilitates the model's ability to integrate visual and textual information during pretraining~\cite{Pix2Struct2023}. The model is trained on 80 million screenshots, enabling it to generalize across diverse domains including the task of widget-captioning of mobile UIs. Pix2Struct achieves significant improvements in the widget captioning task compared to similar approaches~\cite{li2021vut,Donut2022}.

As another recent well-known example, the Pali series of vision-language models uses a vision-based approach to analyze the structural components of UIs~\cite{PaLI-3, PaLI-X}. These models are adept at (among others) visually-situated text understanding, which is crucial for comprehensive widget captioning tasks. To efficiently handle noisy image-text data, they make use of robust pretraining methods, such as contrastive pretraining using SigLIP. 
PaliGemma extends this capability by integrating the SigLIP vision encoder with the Gemma language model, optimizing performance across diverse visual and linguistic tasks~\cite{PaliGemma}. 

Though the deep learning approaches use partial icon context, the recent visual language models (such as Pix2Struct and PaliGemma) require the complete visual context of the UI (such as a complete screenshot) for optimal performance. Requiring full context information implies certain limitations, especially during the iterative and dynamic phases of UI development. In these early stages, complete and finalized UI designs may not always be available, making it challenging to utilize these models effectively.

\section{Ground Truth Analysis}
\label{sec:dataset}

We split a set of annotated icons into two (non-overlapping) subsets. We use the first subset to fine-tune a (text-to-text) large language model and the second one as ground-truth for comparing various annotation approaches. To closely simulate how a third-party human user would annotate icons, we avoid datasets annotated by the original app developers or via AI models. We thus survey existing datasets, looking for large public UI~element datasets that are mobile app focused and annotated by third-party humans.

Most large existing datasets of app UI elements are either proprietary~\cite{ScreenRecognition2021,Crawls2022} or are open but lack third-party human annotations~\cite{IconSeer2023,AITW2023,AitZ_2024,GUI-World_2024,OpenApp2024,GUICourse_2024,AMEX_2024}. Following are good candidates we do not use. IconNet provides 3rd-party annotations but lumps all long-tail icons (beyond 38 classes) into a single ``other'' class~\cite{IconNet2022}. UICaption automatically distills annotations from how-to manuals, which does not always yield good results~\cite{Lexi2022}.

\begin{table}[h]
    \centering
    \begin{tabular}{lrrrr}
        \toprule
        Ground truth & Train & Valid. & Test & Total \\
        \midrule
        WC20 widgets & 52,178 & 4,559 & 4,548 & 61,285 \\
        WC20 captions&138,342 &12,275 &12,242  &162,859 \\
        Our icons    & 18,176 & 1,503 & 1,635 & 21,314 \\
        Our captions (all)& 48,528 & 4,084 & 4,419 & 57,301 \\
        Our captions (1 at random)& 18,176 & 1,503 & 1,635 & \textbf{21,314} \\
        LabelDroid captions   & 15,595 & 1,759 & 1,879 & 19,233 \\
        \bottomrule
    \end{tabular}
    \caption{WC20 (top) has up to three 3rd-party human captions per widget, of which we randomly select one per icon (middle). LabelDroid has a similar number of captions, which are developer-provided (bottom) 
    }
    \label{tab:WC20}
\end{table}

The closest we found is the Widget Captioning data set~\cite{li2020widget}~(``WC20''). WC20 adds to the foundational Rico dataset~\cite{deka2017rico} 12k~mobile UI screens captured via randomly using mobile apps (for a total of 25k~distinct screens from 7k~apps). Each WC20 screen contains both screenshot and a corresponding view hierarchy (aka a JSON DOM tree). In 22k~screens WC20 annotated a total of 61k~UI elements with up to 3 annotations each by distinct human users, yielding 163k 3rd-party human-produced captions. 

\begin{table}[htbp]
\centering
\begin{tabular}{lrrrc}
    \toprule
    Ground truth & Unique & Top-3 & $\le$4 occ. & 1 occ. \\
    \midrule
    LabelDroid~\cite{labeldroid_dataset} & 3,523 & 52\% & 14\% & 12\% \\
    Our (all captions)~\cite{li2020widget} & 24,556 & 8\% & 41\% & 35\% \\
    Our (1 at random)~\cite{li2020widget} & 11,140 & 8\% & 50\% & 43\% \\
    \bottomrule
\end{tabular}
\caption{Icon label distribution:
Unique labels,
share of the top-3 labels, and
labels occurring $\le$4~times / once.}

\label{tab:dataset_analysis}
\end{table}

\begin{wrapfigure}{r}{.4\linewidth}
\centering
\includegraphics[width=\linewidth]{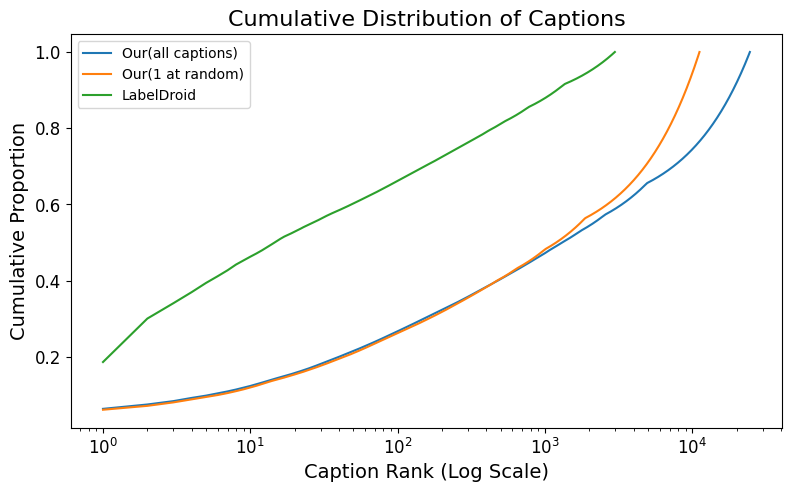}     
\caption{Distribution of icon captions in LabelDroid and our WC20 subset. LabelDroid is more repetitive and less diverse.}
\label{fig:content_distribution}   
\end{wrapfigure}

Of the 22k human-annotated screens 17k are available (via Rico), containing 26k icons (ImageButton or ImageView). We remove abnormally large or narrow elements (as in Coala~\cite{mehralian2021data}), yielding 21,314 3rd-party human annotated icons in 8,201 screens. For each icon we randomly pick one of WC20's annotations, yielding 21,314 annotations. The screen splits we have are consistent with the ones proposed by WC20. Table~\ref{tab:WC20} summarizes the ground truth splits in WC20~\cite{li2020widget}, our work, and LabelDroid~\cite{chen2020unblind}.

While developers often do not provide UI element annotations, we also compare with the diversity of the annotations they do provide. A well-known baseline are the developer-provided UI element labels curated by the LabelDroid work~\cite{chen2020unblind}. While our ``pick-one-WC20-caption-per-icon-at-random'' WC20 subset and LabelDroid have a similar number of total captions (21k vs 19k), Table~\ref{tab:dataset_analysis} and Figure~\ref{fig:content_distribution} show that overall LabelDroid is less diverse than our ground truth, with 3.5k (LabelDroid) vs. 11.1k (ours) unique captions. The three most common captions make up 8\% of our ground truth, whereas 52\% of LabelDroid consists of its top-3 captions.

The diversity in the datasets is also apparent from the Cumulative Distribution Function (CDF) plot shown in Figure~\ref{fig:content_distribution}. Here, the x-axis represents the rank of captions when sorted by frequency, with lower ranks corresponding to more frequent captions and the y-axis represents the cumulative proportion of the total caption occurrences up to that rank. The CDF plot shows that LabelDroid captions have a higher concentration of frequent captions, with a small number accounting for a large portion of the dataset. In contrast, both "Our (all captions)" and "Our (1 at random)" datasets exhibit greater diversity, with more captions needed to cover the same cumulative proportion, indicating less bias and a more evenly distributed set of captions.

\section{Our Approach}

\begin{wrapfigure}{r}{.5\linewidth}
\centering
\includegraphics[width=\linewidth,trim={.5in 3in 5in 2.4in},clip]{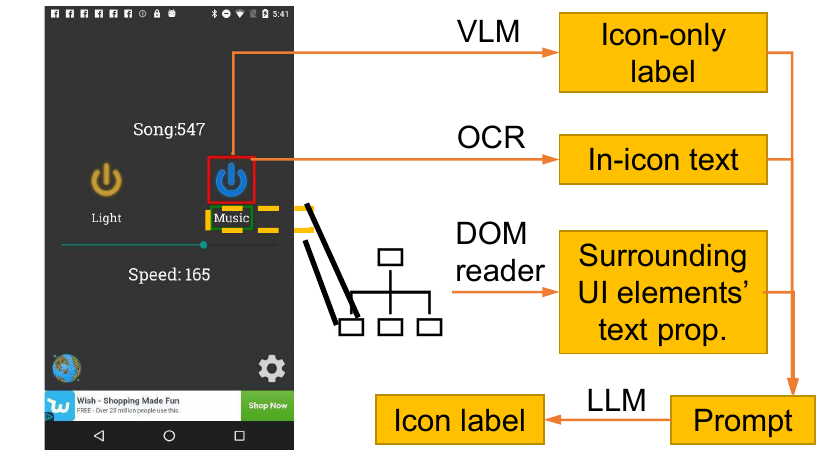} 
\caption{Overview: Via the DOM tree \toolName{} extracts text properties of an icon's surrounding elements (yellow-dashed box), maps the icon's pixels to an icon-only label, and extracts in-icon text.
On this Strobe Light app screen (from Rico) baselines infer
``select the \texttt{<UNK>}'' (Coala) and
``power button'' (PaliGemma).
\toolName{} infers ``turn on the music'' (WC20 ground-truth labels: ``toggle music'' and ``turn on the music'').}
\label{fig:workflow}
\end{wrapfigure}

Figure~\ref{fig:workflow} gives an overview of our approach. \toolName{}'s inputs are a UI screenshot and a view hierarchy (aka DOM tree). Both of these inputs may be under development and hence they may be incomplete, e.g., the screenshot may missing the final shape of other UI elements and the DOM tree may miss several sub-trees.

Via the DOM tree \toolName{} first extracts the current \textbf{icon context} (i.e., the available textual properties of the icon's surrounding UI elements) and the \textbf{icon's pixels}. While allowing fine-tuning with text inputs, at the time of writing none of the OpenAI GPT models nor Gemini allow vision fine-tuning. When using fine-tuning, these models thus only accept text input. 

\toolName{} therefore first passes the icon's pixels to a Large multi-modal Model(LM3) (to map the icon's visual features to a \textbf{icon-only label}). \toolName{} similarly extracts from the icon's pixels any \textbf{in-icon text} via OCR. \toolName{} combines these three kinds of texts into a LLM \textbf{prompt}, yielding an alt-text for the icon.

\subsection{Icon-only Label via LM3}
\label{sec:icon-only-label}

\toolName{} first rips an icon out of its (screen) context and infers a description (or label) for the icon from just the icon's pixels. During app development this could mean just using the icon's file. For our experiments, we cropped each icon from its screenshot via our dataset's DOM tree bounding boxes.

In this work we focus on (clickable) icons via common heuristics. First, we focus on the UI classes containing the substrings IMAGEBUTTON and IMAGEVIEW, as they are most commonly used for clickable icons in Android. During our ground truth selection (and as in Coala~\cite{mehralian2021data}), we skip abnormally large or narrow UI~elements, since they are likely not clickable icons. For our dataset, we thereby remove 17\% of candidate icons for abnormal shapes, yielding the Table~\ref{tab:WC20} icon counts. 

These heuristics may remove some clickable icons (from other UI classes or for odd shapes) and fail to remove clickable images that users may not consider to be a clickable icon. We manually reviewed a random sample of 100 of the oddly shaped (large/narrow) 4,340~icons we removed, finding that 40~of them could reasonably appear to be a clickable icon. This means we ignore some 7\% of total candidate icons (for both training and evaluation) to make our results easier to compare with Coala's evaluation~\cite{mehralian2021data}.

We similarly manually reviewed a random sample of 100 of the resulting 21,314 Table~\ref{tab:WC20} icons, finding (within the context of their screen) 8/100 rather appear to be a clickable image than a clickable icon. We similarly keep these images (both for training and evaluation) to enable easier comparison with Coala's results.

Cropping from screenshots yields icons of various resolutions. As a low resolution may hurt proper icon labeling, we apply Super-Resolution (i.e., Real-ESRGAN~\cite{wang2021real}), yielding a fixed 128x128 pixel resolution.
In an early version we inferred \toolName{}'s icon-only labels with deep learning, i.e., we trained an EfficientNet model~\cite{tan2019efficientnet} on a balanced 100-class icon dataset we curated from Flaticon~\cite{flaticon}, the Noun Project~\cite{noun_project}, and LabelDroid~\cite{chen2020unblind}. While this classifier achieved a 98\% training accuracy and a 94.4\% testing accuracy it was not competitive with the latest Large multi-modal models (LM3s), i.e., Gemini. (Whereas our deep learning classifier used 99 classes + 1 ``other'' class, Gemini produces meaningful descriptions even for many of the long-tail (``other'') icons.)

The current \toolName{} version thus applies Super-Resolution to the icon and then sends the icon to the Gemini LM3 with the following prompt: ``You are an image classifier. What is the class of this UI icon? Only provide the class as response.'' In most cases Gemini's answer consists of a one-word or two-word phrase, which we use as the icon-only label.

\subsection{In-icon Text via OCR}
\label{sec:text_detection}

\begin{wrapfigure}{R}{.25\linewidth}
     \centering
    \begin{minipage}[b]{0.35\linewidth}
        \centering
        \includegraphics[width=\linewidth]{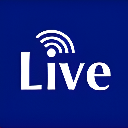}        
    \end{minipage}
    \hfill
    \begin{minipage}[b]{0.5\linewidth}
        \centering
        \includegraphics[width=\linewidth]{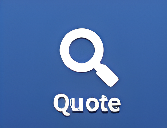}       
    \end{minipage}
    \caption{In-icon text examples from Rico.}
    \label{fig:ocr}   
\end{wrapfigure}

Some icons contain text that may provide information about the icon's purpose. Such text is however usually not directly available in the DOM tree (e.g., via the icon's ``text'' property). We thus heuristically retrieve such in-icon text, via Optical Character Recognition (OCR) on the (cropped) icon pixels, specifically using the latest version (1.7.1) of EasyOCR~\cite{easy_ocr}. 

Figure~\ref{fig:ocr} shows two examples icons containing text that would be helpful for alt-text generation. For the Figure~\ref{fig:ocr} examples \toolName{}'s OCR correctly inferred the in-icon text strings ``Live'' and ``Quote''.

\subsection{Surrounding UI Elements' Textual Properties From the DOM Tree}
\label{sec:icon-context}

\begin{wrapfigure}{r}{.45\linewidth}
\begin{lstlisting}[caption={Icon context \toolName{} extracts from the DOM tree \& OCR for the Figure~\ref{fig:workflow} example.}, label={lst:icon_context}, breaklines=true,numbers=none]
{
  "app activity name": "com.ape.apps.strobe.StrobeActivity",
  "UI element info": {
    "class_name": "AppCompatImageView",
    "resource_id": "ivTechnoPower",
    "bounds": [957, 878, 1202, 1123],
    "OCR detected text": ['d)']
  },
  "parent node": {
    "class": "LinearLayout"
  },
  "sibling nodes": [
    {
      "resource_id": "tvTechno",
      "text": "Music",
      "class": "AppCompatTextView"
    }
  ]
}
\end{lstlisting}
\end{wrapfigure}

To provide an icon's context in a LLM query, previous works~\cite{wang2023enabling,wen2024AutoDroid} encoded a screen's entire view hierarchy. While this captures a lot of context information, this approach has two challenges. First, many mobile screens have a complex hierarchy, leading to large queries, which may be more expensive to process (for both fine-tuning and inference). Second, encoding an entire DOM tree may not be possible, as during software development the full tree may not yet be available.

We conducted initial experiments with zero-shot prompting, comparing full DOM-tree context with a small context subset (Table~\ref{tab:zero-shot}). Since we observed similar performance, we pursue a more limited context and support partial DOM trees during software development.

Listing~\ref{lst:icon_context} shows an example icon context \toolName{} extracts for the Figure~\ref{fig:workflow} motivating example from the DOM tree. The first component is the screen's name (``StrobeActivity''), followed by properties of the icon (class name, resource-id, bounding box location, and on-screen text). The Figure~\ref{fig:workflow} example icon did not come with a on-screen text property (and our resulting DOM extract just omits any such missing elements). At this point we also add any in-icon text \toolName{} inferred via OCR (i.e.: ``d)'').

Besides screen name and the icon, we extract similar information (class, resource-id, and on-screen text) about the icon's parent node in the view hierarchy as well as the icon's sibling nodes (sharing the icon's parent node). Our intuition is that the developer grouping these UI elements together into the same container likely captures that they are conceptually related, yielding valuable context information.

Throughout this DOM extraction, if a property such as an on-screen text is missing or blank in the DOM tree then we do not add a corresponding key-value pair to our DOM-extract. \toolName{} currently also does not extract any alt-texts that may already be in the screen's DOM tree but we could easily add this in a future version (which would likely improve \toolName{}'s inferred alt-texts).

\subsection{Prompt Generation for Alt-text Labeling}

\begin{wrapfigure}{r}{.7\linewidth}
\begin{lstlisting}[caption={Prompt template \toolName{} uses for its LLM queries.}, label={lst:prompt},numbers=none]
"You are an accessibility assistant to a mobile app Developer. 
A mobile app UI element that looks like an icon that developers 
often use with icon tag '{icon-only label}' has view hierarchy 
content as below: 
{icon context}
Generate a short (within 2-7 words), descriptive alt-text for the 
UI element. Provide only the alt-text as output, nothing else. 
Describe the element as if you were the app developer to help 
vision-impaired user understand its functionality and purpose. 
Avoid generic words like 'button', 'image', 'icon' etc."
\end{lstlisting}
\end{wrapfigure}

\toolName{} wraps its icon-only label and icon context in a LLM query template. We crafted this template to convey \toolName{}'s goals and briefly describe \toolName{}'s inferred text to the LLM in natural language. 

Listing~\ref{lst:prompt} shows the the template we used to evaluate \toolName{}. \toolName{} replaces `\{icon-only label\}' with the label it inferred from the icon pixels via a LM3(Section~\ref{sec:icon-only-label}) and \{icon context\} with the sub-tree \toolName{} inferred from the screen's DOM tree and OCR (Section~\ref{sec:icon-context}).

\section{Quantitative \& Qualitative Evaluation}

In our evaluation of alt-text generation for UI icons, we assessed the performance of our approach using off-the-shelf large language models (LLMs) and compared these results against both traditional deep-learning models, vision-transformers, and advanced vision-language models (VLMs). Specifically, we compare with deep-learning models such as Coala~\cite{mehralian2021data} and LabelDroid~\cite{chen2020unblind}, and newer approaches such as the vision transformer based Pix2Struct~\cite{Pix2Struct2023} and the PaliGemma VLM~\cite{PaliGemma}. 

While PaliGemma is a versatile VLM that integrates visual and linguistic data for enhanced text generation, Pix2Struct is a model focused on parsing visual data from screenshots and generating corresponding textual descriptions. Both PaliGemma and Pix2Struct are fine-tuned on full UI screen images for the widget-captioning task, with each model demonstrating its strengths in handling visual and contextual information within user interfaces.

\begin{wrapfigure}{R}{.5\linewidth}
     \centering
    \begin{minipage}[b]{0.48\linewidth}
        \centering
        \includegraphics[width=\linewidth]{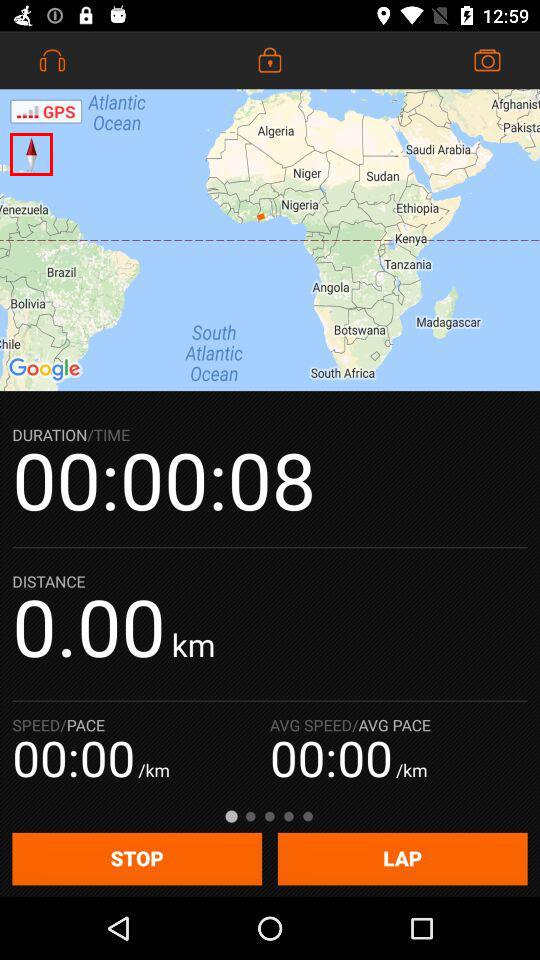}        
    \end{minipage}
    \hfill
    \begin{minipage}[b]{0.48\linewidth}
        \centering
        \includegraphics[width=\linewidth]{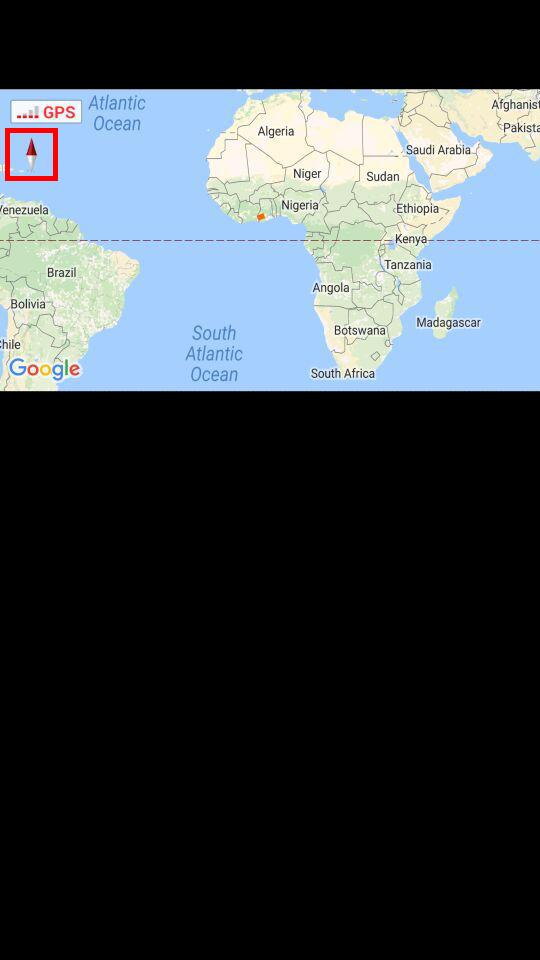}       
    \end{minipage}
    \caption{Input formats for Pix2Struct and PaliGemma}
    \label{fig:PaliGemma_input}   
\end{wrapfigure}

For a more targeted comparison, we considered two input configurations for the modern approaches (vision-transformers and VLMs): full-screen inputs and masked inputs-which contain only the icon and its neighboring elements(Figure ~\ref{fig:PaliGemma_input}. This approach aligns with our goal to evaluate performance under conditions where the complete UI screen might not be available, such as during the early stages of app development.

We seek to answer the following research questions (LM3 refers to a large multi-modal model).
\begin{description}
    \item[RQ0] For icon alt-text generation, how do traditional deep-learning approaches compare with large modern out-of-the-box multi-modal models and \toolName{} (i.e., without fine-tuning)?
    \item[RQ1] For icon alt-text generation, can a vision-then-text LM3-LLM
     pipeline approach the results of the LM3?
    \item[RQ2] For icon alt-text generation with a LM3-LLM pipeline, can a local icon context yield similar results as full screen context?
    \item[RQ3] For icon alt-text generation, how do modern fine-tuned vision-transformers and VLMs compare with the \toolName{} LM3-LLM pipeline?
    \item[RQ4] For icon alt-text generation, how well do modern fine-tuned models support partial input screens?
    \item[RQ5] How do users rate the icon alt-texts of the models scoring highest on automated metrics?
\end{description}

\subsection{Experimental Setup}

We empirically compare \toolName{} with representative tools from the most-closely related approaches, i.e., Coala \& LabelDroid for traditional deep-learning approaches, Pix2Struct for the recent vision-transformers, and PaliGemma for VLMs. To make this comparison as fair as possible, we evaluated all tools on the same 1,635~WC20-derived test icons (Table~\ref{tab:WC20} row ``our captions (1 at random)''). We similarly trained and fine-tuned all tools on the same dataset, i.e., WC20~\cite{li2020widget}.

Specifically, we downloaded the LabelDroid code\footnote{\url{https://github.com/chenjshnn/LabelDroid}, accessed Sept. 2024} and trained it on the LabelDroid training data we obtained from Zenodo. This allowed us to reproduce the results reported in the LabelDroid paper. We then further trained this LabelDroid model on our icon dataset, i.e., using the icon-label pairs in the exact split described in the Widget Captioning paper (the Table~\ref{tab:WC20} row labeled
``our captions (1 at random)''). 

Similarly, we acquired the Coala code\footnote{\url{https://github.com/fmehralian/COALA}, accessed Sept. 2024} and replicated their results using the default configuration on their UI30K dataset The replication results for both DL approaches can be found here~\cite{our_code}. Similar to our work, Coala utilizes both the icon image and its usage context from the view hierarchy. Since the view hierarchy information in our data(the Table~\ref{tab:WC20} row labeled
``our captions (1 at random)'') is in JSON format and Coala's code required them in XML, we first converted our data to XML. We also rearranged them in the format required by Coala- i.e. into folders based on the app package. We later trained their context-aware model without attention - as reported in their paper to achieve the best result on this dataset and obtained the scores in Table~\ref{tab:finetuning_result}. Both the DL models report higher image captioning scores on their dataset than ours- possibly because of the generic nature of the icon labels in their dataset to the more informative and diverse dataset we obtained from WC20~\cite{li2020widget}. These were trained on a 16G RAM intel core i7-11700F machine with NVIDIA GeForce RTX 3060 Ti GPU.

For the vision-transformer and VLM approaches, we utilized models from HuggingFace that were already fine-tuned on the WC20 dataset, specifically Pix2Struct\footnote{\url{https://huggingface.co/google/pix2struct-widget-captioning-large}, accessed Sept. 2024} and PaliGemma\footnote{\url{https://huggingface.co/google/paligemma-3b-ft-widgetcap-448}, accessed Sept. 2024}. Both models were trained on the full WC20 dataset, which includes various UI elements, not just icons. As a result, the results reported in this paper may differ from those in the original papers for these models.

To fine-tune the large language model (LLM) for our tool, \toolName{} on diverse icons, we took the 99 common icon classes Liu et al.~\cite{liu2018learning} identified in Rico, added a class ``other'', and sampled up to 15 icons per class from our 18,176~training icons, yielding 1,425 icons. For fine-tuning we provided the LLM with WC20's human-inferred label and \toolName{}-inferred icon-only label and icon context. For zero-shot prompting and LLM fine-tuning we used the OpenAI~\cite{openai} and Gemini~\cite{gemini_api} APIs, staying within a total budget of USD~100.

\subsection{Metrics: Distance to a Set of Human-written Ground-truth Icon Alt-texts}

For an initial automated evaluation we use the 1,635 ground-truth test pairs (icon within a screen plus up to three human-written icon alt-texts) from Table~\ref{tab:WC20}. Following common practice we use the MS COCO~\cite{ms_coco} implementation to compute the following widely-used text distance metrics to capture how well the tool-generated icon alt-text matches the up-to three human-written reference texts. Overall, higher scores across these metrics are better, signifying a closer match between generated texts and the ground-truth data. While widely used in the current research literature, each of these metrics has limitations and thus does not fully encapsulate human judgment.

BLEU~\cite{papineni2002bleu} computes a precision measure on how close a tool-generated candidate sentence is to the up to 3 human-written sentences.
To compute precision, at its core BLEU~\cite{papineni2002bleu} counts exact matches of word sequences (or n-grams). As icon labels are often short, we focus on BLEU-1 and BLEU-2 (which matches 1-grams and 2-grams). Among the limitations, BLEU penalizes spelling differences, synonyms, and paraphrases.  

ROUGE~\cite{lin2004rouge} counts word sequence matches similar to BLEU but focuses on recall, assessing how much information in the reference texts the candidate text captures. While at its core it generalizes from BLEU's continuous-sequence matches to longest common subsequences (e.g., 3 for (A B C) and (A X B Y Z), ROUGE still penalizes spelling differences, synonyms, and paraphrases.

METEOR~\cite{banerjee2005meteor} combines features from both BLEU and ROUGE (including precision and recall). While generalizing from these earlier metrics' exact word-level matching to also include stemming and synonyms based on WordNet~\cite{miller1995wordnet}, METEOR still penalizes phrases that are semantically related but not synonyms. METEOR also generalizes ROUGE's subsequence matching to out-of-order alignments, which may not fully capture the nuances of word order critical to human perception.

CIDEr~\cite{vedantam2015cider} goes beyond n-gram precision by incorporating TF-IDF weighting~\cite{ramos2003using}, assigning higher scores to informative n-grams less frequent in the reference set. It considers overlap with multiple reference captions, reflecting the variability in human descriptions while still capturing the essence of the image. This method prioritizes rare but informative phrases that are significant in the given context, making it highly effective for tasks like ours where the distinctiveness of each description is crucial. However, CIDEr sometimes can result in unbalanced TF-IDF weighting, causing unimportant details to be weighted more, resulting in ineffective caption evaluation~\cite{re-evaluating_captioning_metrics}.

We also integrate SPICE~\cite{anderson2016spice} to address semantic accuracy by examining the agreement of propositions within the texts, providing a deeper assessment that more closely mirrors human evaluations of relevance and accuracy. It complements CIDEr by ensuring the generated descriptions are not only informative but also semantically aligned with human judgments. SPICE builds on scene graphs that parse semantic tokens from candidate and reference sentences, and thus incorrect parsing results may lead to inaccurate evaluation~\cite{re-evaluating_captioning_metrics}. Besides, though scene graphs have proven highly effective for complex image retrieval~\cite{scene_graph}, alt-text to UI icons are often much simpler and their functionality may not always map well to detailed scene graphs.

While each metric offers valuable perspectives, CIDEr and SPICE are most suited for our task of evaluating alt-texts since they align the closest to human judgement~\cite{anderson2016spice}.

\subsection{Gauging Multi-modal Model \& LM3-LLM Capabilities via Zero-shot Prompting}

\begin{table*}[h!t]
    \centering
    \begin{tabular}{lccccccc}
        \toprule
        DL vs. 0-shot large models & & BLEU-1 & BLEU-2 & ROUGE & METEOR & CIDEr & SPICE \\
        \midrule
        Coala (no attention) & & 35.3 & \underline{24.1} & 34.0 & 14.2 & 66.3 & 9.4 \\
        LabelDroid & & 28.4 & 17.7 & 31.7 & 12.5 & 59.4 & 10.6 \\
        \midrule
        Gemini 1.5 Flash (vision) & part & \textbf{44.8} & \textbf{28.0} & \textbf{46.6} & \textbf{22.8} & \textbf{94.4} & \textbf{19.9}\\
        \midrule
        \toolName{} (GPT 4o)  & part & \underline{37.8} & 19.4 & \underline{40.3} & \underline{19.3} & \underline{70.1} & \underline{15.7} \\
        \toolName{} (GPT 3.5) & part & 30.0 & 15.0 & 34.7 & 18.2 & 56.3 & 14.5 \\
        \toolName{} (GPT 3.5) & full & 26.7 & 13.2 & 32.6 & 18.5 & 51.7 & 14.3 \\
        \bottomrule
    \end{tabular}
    \caption{Traditional deep-learning approaches (top) vs zero-shot use (without fine-tuning): multi-modal model (middle) \& LM3-LLM pipelines (bottom);
    part~=~partial icon context;
    full~=~full hierarchy;
    bold~=~best score;
    underlined~=~runner-up.
    With zero-shot prompting, Gemini scores highest, followed by \toolName{} (GPT 4o). Except on METEOR, \toolName{} (GPT 3.5) with partial icon context performs better than with full hierarchy.
    }
    \label{tab:zero-shot}
\end{table*}

\subsubsection{RQ0: Large Modern Models Yield Better Alt-texts Even With Zero-shot Prompting}

Recent advances have produced very large multi-modal models such as recent OpenAI GPT models, Gemini~\cite{Gemini_2024}, and Gemini~1.5~\cite{Gemini-1.5}, which have been trained on (among others) text and images. While these recent multi-modal models allow fine-tuning with text inputs, at the time of writing none of these GPT nor Gemini models allow vision fine-tuning. Given their extensive training, it is still an interesting baseline to compare their performance out-of-the-box (i.e., without any fine-tuning) versus traditional deep-learning approaches that are specialized on this task (i.e., LabelDroid and Coala).

For this experiment, we submitted prompts to Gemini 1.5 Flash (vision) that mimic the prompts \toolName{} generates (Listing~\ref{lst:prompt}). Specifically, the Gemini prompts include the icon context \toolName{} extracts from the DOM tree (including the OCR-inferred in-icon text) but replaces the LM3-inferred icon-only label with the pixels of the cropped icon. The results in Table~\ref{tab:zero-shot} show how much better this zero-shot use of Gemini performed vs. the traditional specialized deep-learning approaches LabelDroid and Coala, especially on the most relevant metrics CIDEr and SPICE.

At a high level, we conclude that for our task our custom use of a multi-modal model is likely to outperform a specialized deep-learning approach (as observed in Table~\ref{tab:zero-shot}). We further conclude that this outperformance is likely to become even larger once fine-tuning becomes available for large multi-modal models (as fine-tuning large models has improved performance in the past).

\subsubsection{RQ1: A LM3-LLM Pipeline Can Produce Similar Results to a Multi-Modal Model}

We do not know if or when multi-modal fine-tuning becomes available and how expensive it may be. In the meantime we thus work with large models that allow fine-tuning, i.e., based on input text. Based on our Table~\ref{tab:zero-shot} experiment, we conclude that a promising approach to describe an icon image is to use a large multi-modal model (vs. training a deep-learning model). This aligns with the intuition that a large multi-modal model has been trained on a much larger set of images, yielding better image understanding and more nuanced image descriptions.

As a baseline we thus compare a large zero-shot multi-modal model (LM3) with a zero-shot LLM that in a pre-processing (phase 1) step describes the icon image with the same powerful zero-shot LM3. Such a pipelined setup then allows us to fine-tune the (phase 2) LLM component.

For this experiment we used different versions of \toolName{} in a zero-shot configuration (without any fine-tuning). First, we experimented with both GPT~3.5 and the more recent GPT~4o. Table~\ref{tab:zero-shot} shows that, as expected, just switching \toolName{} from GPT~3.5 to GPT~4o improves results (across all metrics). It is notable that the GPT~4o version gets relatively close to the multi-modal Gemini results, despite the GPT~4o version only operating on text input.

\subsubsection{RQ2: Local Icon Context Can Yield Similar Results as Full Screen Context}

Finally, we are curious if providing a larger amount of context improves \toolName{}'s results. We thus compare \toolName{} with a version that replaces \toolName{}'s DOM-extracted icon context with the screen's full DOM tree (in condensed HTML format). Table~\ref{tab:zero-shot} results show that (except for a minor improvement in the METEOR score) overall this additional context did not improve results, maybe burying the most relevant context information in a large amount of noise.

\subsection{Comparing with State of the Art Approaches Including Fine-tuning}

\begin{table*}[h!t]
\centering
\begin{tabular}{lrrrrrr}
\toprule
Model & BLEU-1 & BLEU-2 & ROUGE & METEOR & CIDEr & SPICE \\
\midrule
G-1.5 (0-shot, partial context) & 44.8 & 28.0 & 46.6 & 22.8 & 94.4 & 19.9\\      \toolName{} (0-shot, partial context) & 30.0 & 15.0 & 34.7 & 18.2 & 56.3 & 14.5 \\  
\midrule
Pix2Struct (large) (full screen) & 41.9 & 28.1 & 41.7 & 18.4 & 89.5 & 14.8 \\
Pix2Struct (large) (masked screen) & 35.6 & 23.9 & 35.1 & 15.7 & 73.7 & 10.7 \\
PaliGemma (full screen) & \textbf{56.5} & \textbf{42.1} & \textbf{56.5} & \underline{25.1} & \underline{127.9} & \underline{21.0} \\
PaliGemma (masked screen) & 52.4 & 38.6 & \underline{55.0} & 23.6 & 121.5 & 19.6 \\        

\toolName{} (partial context) & \underline{55.2} & \underline{39.0} & 54.2 & \textbf{26.1} & \textbf{129.6} & \textbf{21.9} \\
\bottomrule
\end{tabular}
\caption{Zero-shot large models (top, from Table~\ref{tab:zero-shot}) vs. vision-transformer, VLM, and \toolName{} (all fine-tuned, bottom).
\toolName{} scores highest (bold) in the most-relevant metrics CIDEr \& SPICE. The runner-up (underlined) needs full-screen inputs.}
\label{tab:finetuning_result}
\end{table*}

While Table~\ref{tab:zero-shot} has shown better results on GPT~4o, for \toolName{} we picked GPT~3.5 to ensure that fine-tuning remains within our financial budget. Specifically, for \toolName{} we fine-tuned OpenAI's gpt-3.5-turbo-0125 model~\cite{openai_finetuning} for 3 epochs on our 1,425~icons (sampled from our 18,176~training icons). On the positive side, it should be straight-forward to improve \toolName{}'s results by replacing \toolName{}'s fine-tuned GPT~3.5 with GPT~4o fine-tuned on our 1,425~icons.

\subsubsection{RQ3: \toolName{} Outperforms SOTA Models in CIDEr \& SPICE}

We compare \toolName{} with the state-of-the-art (SOTA) approaches Pix2Struct and PaliGemma. These SOTA approaches are fine-tuned on the entire WC20 dataset. (WC20 is a superset of our training set, which in turn is a superset of the 1,425 icons \toolName{} uses for fine-tuning.) We follow their experimental setup~\cite{Pix2Struct2023,PaliGemma}, by providing Pix2Struct and PaliGemma as input a screenshot where the target icon is marked with a red bounding box. To capture how well they perform on partial screens that may occur during software development, we ran these SOTA approaches both on the entire screenshot and on a masked screen that blacks out everything except for the icon and its sibling UI elements.

Table~\ref{tab:finetuning_result} first shows that the (fine-tuned) \toolName{} outperforms our Table~\ref{tab:zero-shot} zero-shot experiments (which in turn outperformed traditional deep-learning approaches). As expected, Table~\ref{tab:finetuning_result} shows fine-tuning has a big impact on \toolName{} (e.g., increasing CIDEr scores from 56.3 to 129.6). Table~\ref{tab:finetuning_result} further shows that \toolName{} also outperforms SOTA models in the most-relevant metrics (CIDEr and SPICE) in all configurations.

\subsubsection{RQ4: \toolName{}'s Improvement on SOTA is Larger for Partial Screens}

This outperformance becomes larger when comparing with the SOTA on the partial screens that may be found in our target task during software development.
Following are concrete examples to illustrate differences between \toolName{} and the next-highest scoring approaches, i.e., PaliGemma full and PaliGemma masked.

\begin{figure*}[h!t]
\centering
\subcaptionbox{``search for numbers'', ``search phone number'' (ref); 
``search'' (P\textsubscript{f} \& P\textsubscript{m});
``search for number'' (ours).
\label{fig:screen_search_1B_numbers}}{
    \includegraphics[width=0.31\textwidth]{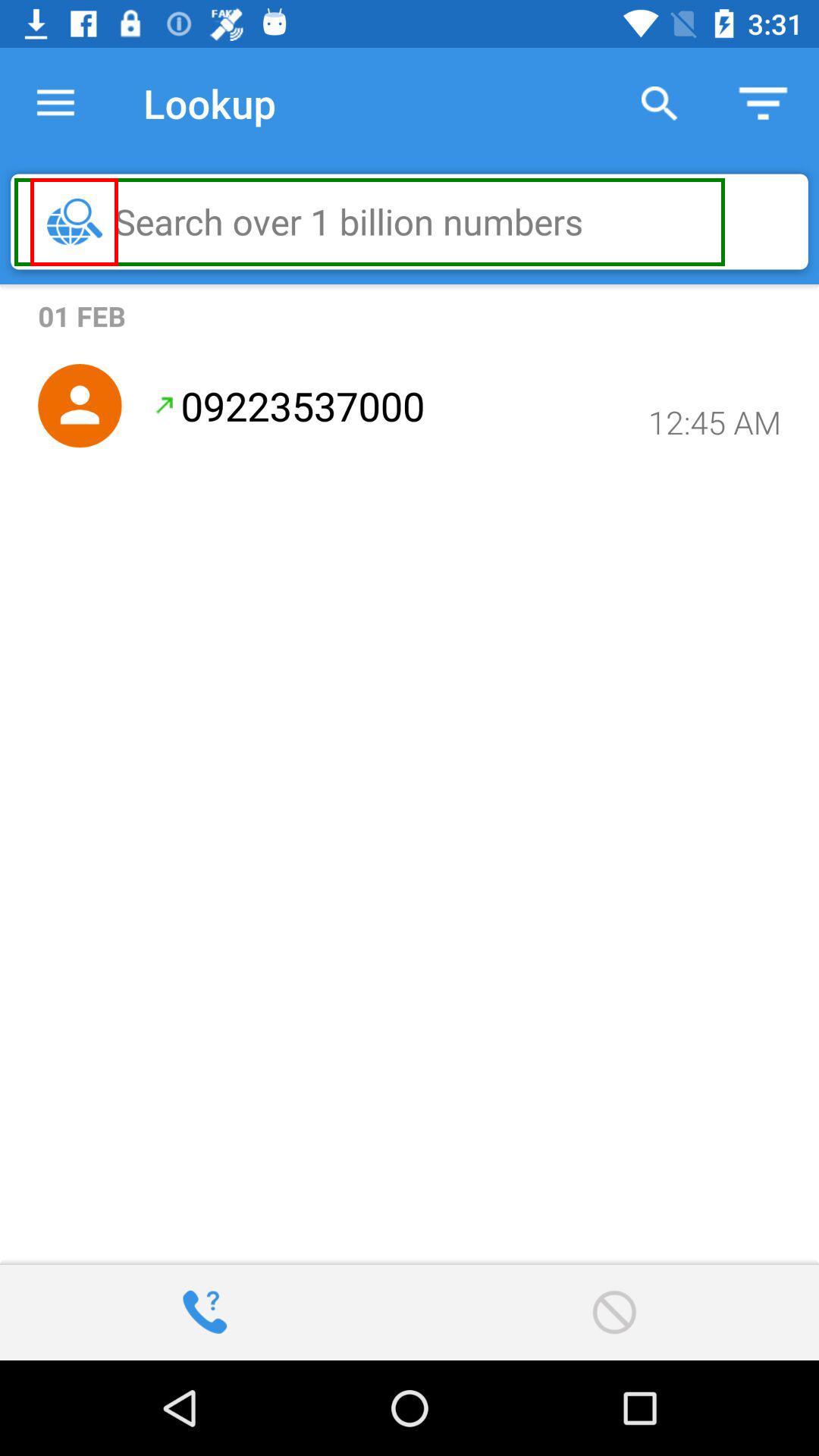}
}
\hfill
\subcaptionbox{``cancel'', ``clear out phone number'', ``delete input'' (ref);  
``close'' (P\textsubscript{f} \& P\textsubscript{m});
``clear search box'' (ours).
\label{fig:screen_clear_phone_nr}}{
    \includegraphics[width=0.31\textwidth]{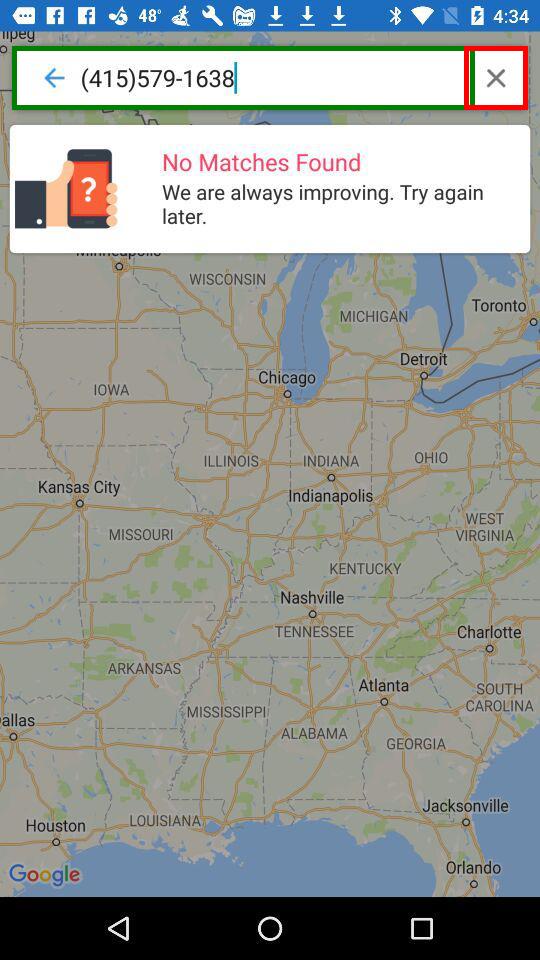}
}
\hfill
\subcaptionbox{``submit name button'', ``the name has been approved'' (ref); 
``select name'' (P\textsubscript{f});
``open menu'' (P\textsubscript{m});
``confirm name'' (ours).
\label{fig:screen_raphael_tan}}{
    \includegraphics[width=0.31\textwidth]{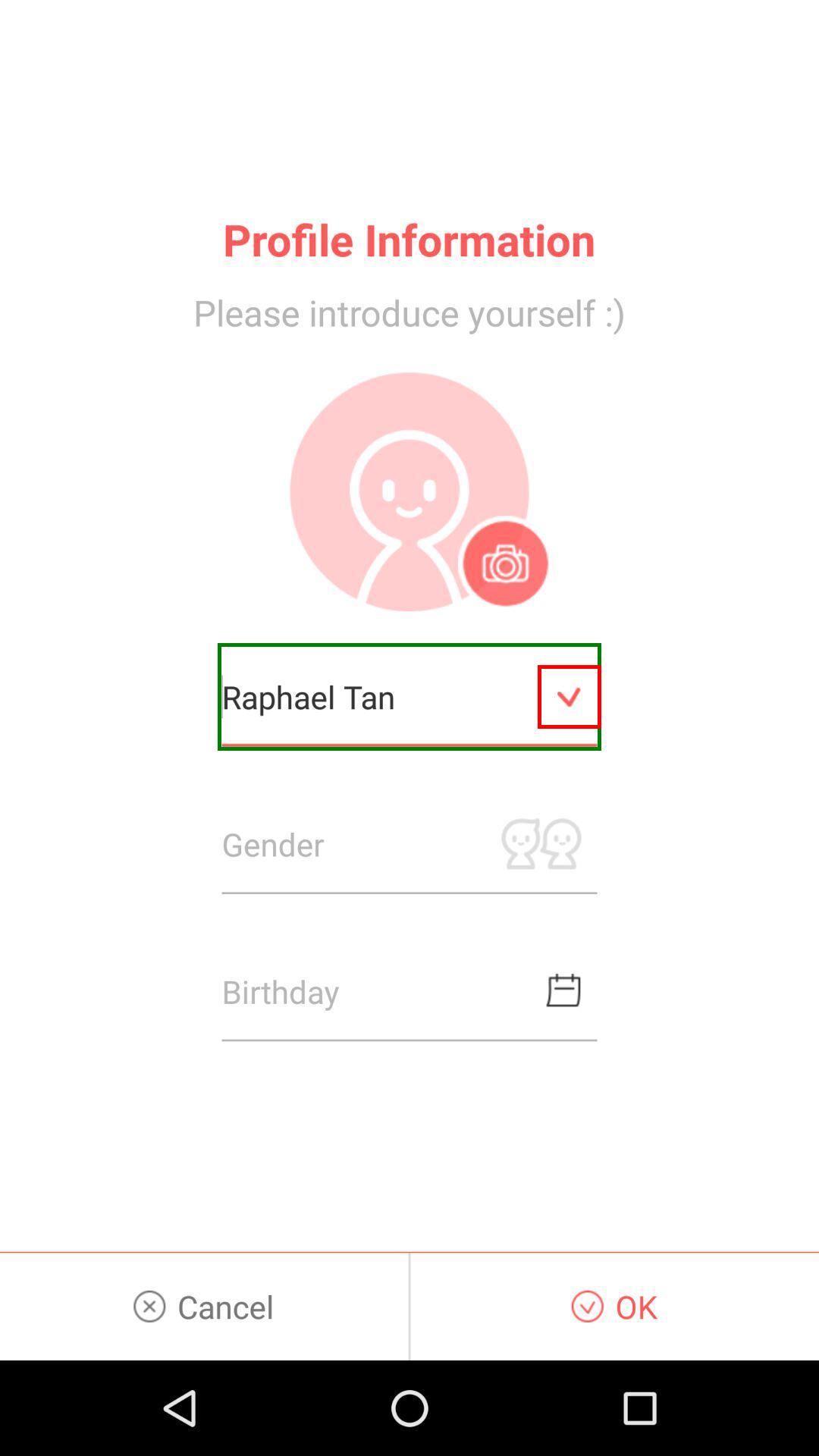}
}
\caption{Sample screens with icons marked via red box and alt-texts inferred by humans and highest-scoring models;
green box~=~icon context (green markup not part of tool image input);
ref~=~ground truth;
P\textsubscript{f}~=~PaliGemma full;
P\textsubscript{m}~=~PaliGemma masked;
ours~=~\toolName{}.
}
\label{fig:good_examples}
\end{figure*}

\subsubsection{Explaining Icons With A Sibling}

The icon in Figure~\ref{fig:screen_search_1B_numbers} combines the loop symbol commonly used for search with a globe symbol. Both PaliGemma versions seem to ignore the icon context and produce ``search''. Within \toolName{}, Gemini similarly infers ``GlobalSearch''. While the \toolName{}-inferred DOM context is noisy with several generic identifiers, it also contains the sibling search box string ``Search over 1 billion numbers'', which allows \toolName{} to infer ``search for number''.

\subsubsection{Explaining Common Icon Shapes With Local DOM Context}

The Figure~\ref{fig:screen_clear_phone_nr} example is interesting because the target \code{[x]} icon looks like a common ``close'' button. The purely-visual PaliGemma struggled to combine this information with the icon's context, just inferring ``close'' (both on the full and masked screen). Within \toolName{}, Gemini similarly infers icon-only label ``CloseButton''.

On the other hand, the human ground truth (``cancel'', ``clear out phone number'', ``delete input'') aligns with the icon's DOM context, which contains (among others) the term ``search'' five times. One of these is in the icon's resource id (\code{search\_card\_\_clearTextButton}), which also contains ``clear''. The context further contains ``edit'' four times and ``text'' five times. Taken together these make it easier for \toolName{} to infer the icon label ``clear search box''.

\subsubsection{Explaining Icons With Limited Context Information}

The Figure~\ref{fig:screen_raphael_tan} icon looks like a check-mark and \toolName{}'s Gemini infers ``Checkmark''. On the full screen PaliGemma infers a reasonable ``select name''. But when only run on the icon siblings (masked) PaliGemma infers ``open menu'', likely because many drop-down menus use a similar check-mark style.

While \toolName{}'s DOM-tree extract only contains a few context elements, it contains the resource ids of both the icon (\code{nickname\_check}) and its sibling (\code{nickname\_input}). Together this was enough for \toolName{} to infer ``confirm name'', which is similar to the ground truth (``submit name button'', ``the name has been approved'').

\subsection{For Partial Screens Users Rate \toolName{}'s Icon Alt-texts Highest (RQ5)}

To complement our automated metric-based evaluation, we also conducted a small user study. The goal of the study is to simulate how useful an app user would rate a given icon annotation. As smartphones are ubiquitous we sampled participants from our local student population. All 10 participants were adults (7~male, 3~female). Participation was voluntary and unpaid.

For this study we randomly select from our test set 50 screen/icon pairs. For each screen we mark the target icon via a bounding box and present a randomized list of four alt-texts (PaliGemma full and masked, ours, and a randomly-selected ground truth), all in a Google form. Participants were instructed to assess each alt-text for its accuracy in describing the icon’s functionality and purpose within its UI context, by rating them on a scale of 1 (worst) to 5 (best). This prompt aims to determine if an alt-texts is sufficient for users with visual impairments. 

Each response was complete, rating 4 annotations per screen, yielding a total of 2k~annotation ratings. On average, participants rated the alt-text
of the ground truth~3.6,
full-screen PaliGemma~3.8,
masked PaliGemma~3.4, and
\toolName{}~3.8. 
These results are consistent with our automated metric study in that (1)~full-screen PaliGemma is rated similarly as \toolName{} and (2)~both approaches rate higher than masked PaliGemma.

Rating variability, expressed via standard deviation~(SD), was highest for
masked PaliGemma and ground truth (SD = 1.5), indicating diverse perceptions among the participants. Conversely, \toolName{} demonstrated the lowest variability (SD = 1.3), suggesting a more consistent reception among the evaluators.

\begin{table}[h!t]
\centering
\begin{tabular}{lcc}
\toprule
\textbf{Comparison} & \textbf{p-value} & \textbf{Reject Null Hyp.} \\
\midrule
\toolName{} vs PaliGemma-full & 0.6700 & No \\
\toolName{} vs PaliGemma-masked & $3.54 \times 10^{-8}$ & Yes \\
\bottomrule
\end{tabular}
\caption{Hypothesis testing results using the Wilcoxon signed-rank test.}
\label{tab:wilcoxon_rank}
\end{table}

Given the failure of the Shapiro-Wilk test for normality~\cite{Shapiro-wilk}, we opted for the Wilcoxon signed-rank test~\cite{wilcoxonranktest}, a non-parametric method suitable for our samples. We used this test to statistically assess the differences between the quality of alt-text generated by \toolName{} and the other methods, with a 0.05~significance threshold. The Table~\ref{tab:wilcoxon_rank} results show no significant difference between \toolName{} and full-screen PaliGemma. On the other hand, there is a statistically significant difference between \toolName{} and masked PaliGemma, indicating \toolName{}'s superior performance in generating contextually appropriate alt-texts for partial screens.

\subsection{Causes of Error}

\begin{figure*}[h!t]
    \centering
    \subcaptionbox{``dislike comment'', ``vote against this comment'', ``vote down'' (ref);
    ``download option'' (ours).
    \label{fig:class_effect}}{
        \includegraphics[width=0.23\textwidth]{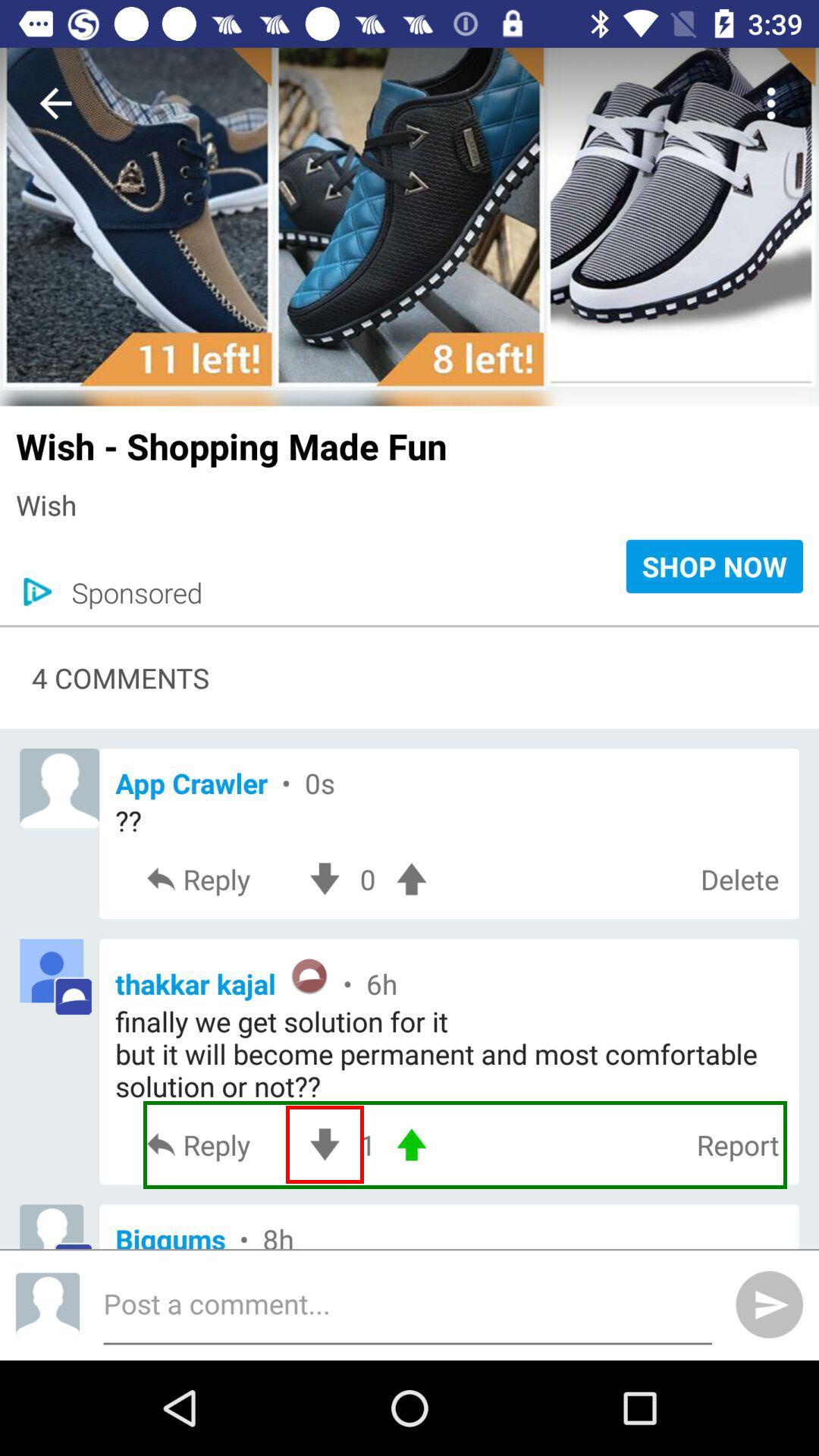}
    }
    \hfill
    \subcaptionbox{``dollars'', ``enter funds or payment option'', ``show total balance'' (ref);
    ``buy hint'' (ours).
    \label{fig:context_effect}}{
        \includegraphics[width=0.23\textwidth]{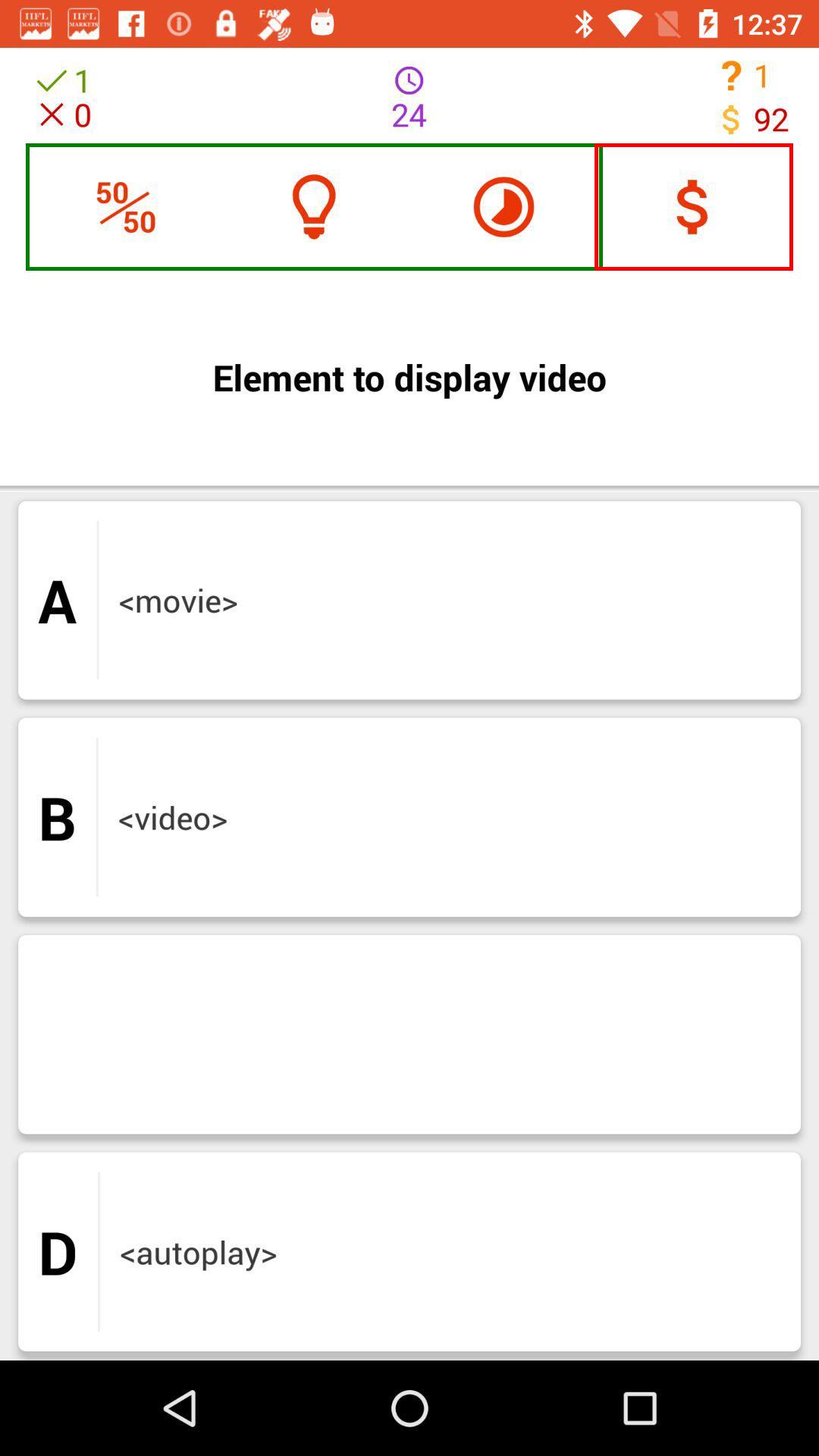}
    }
    \hfill
    \subcaptionbox{``escuchar musica'', ``select human'', ``specific image selection'' (ref);
    ``error icon'' (ours).
    \label{fig:not_icon}}{
        \includegraphics[width=0.23\textwidth]{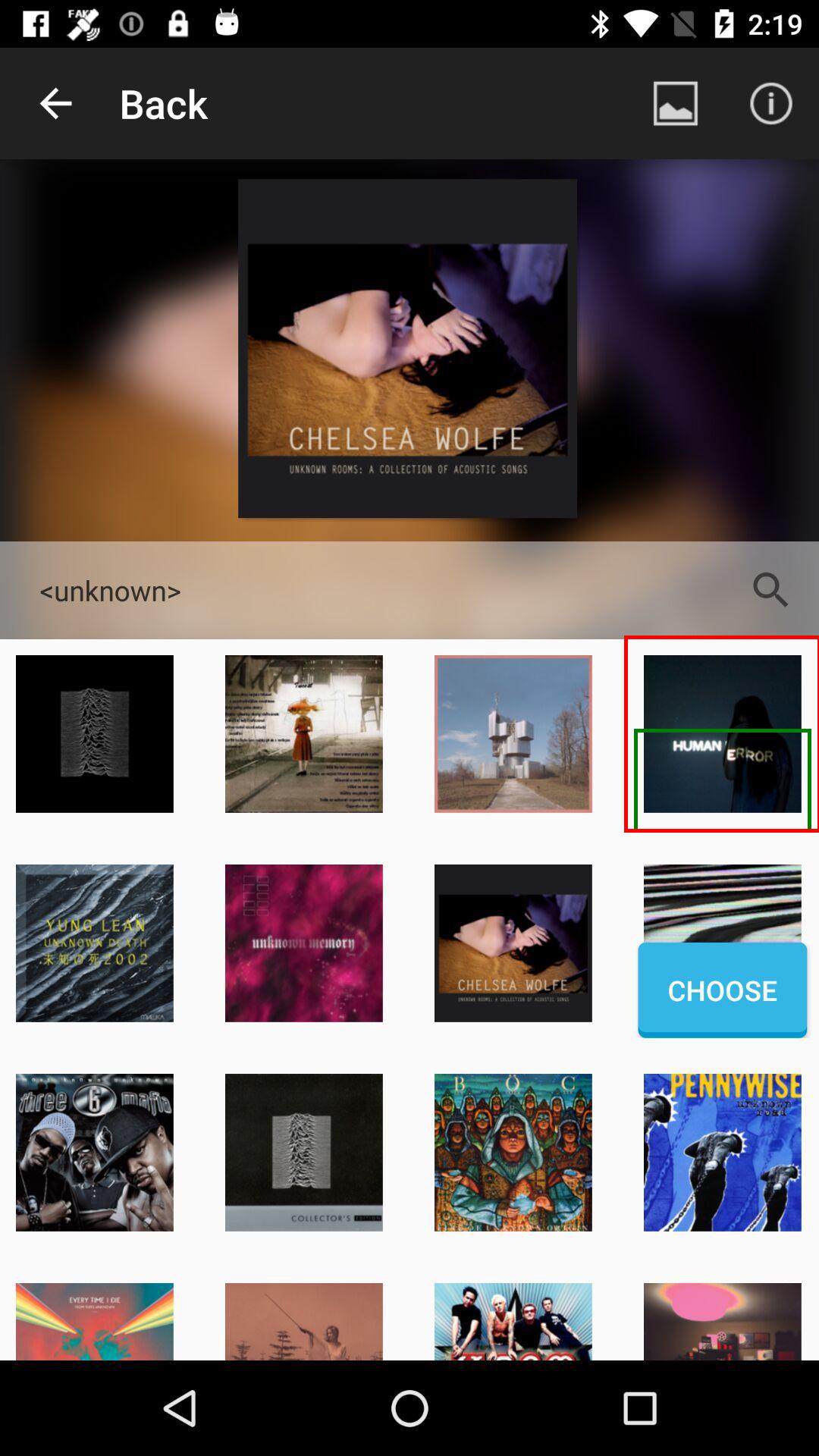}
    }
    \hfill
    \subcaptionbox{ref see below;
    ``decrease the quantity of the item'' (ours).
    \label{fig:confusing}}{
        \includegraphics[width=0.23\textwidth]{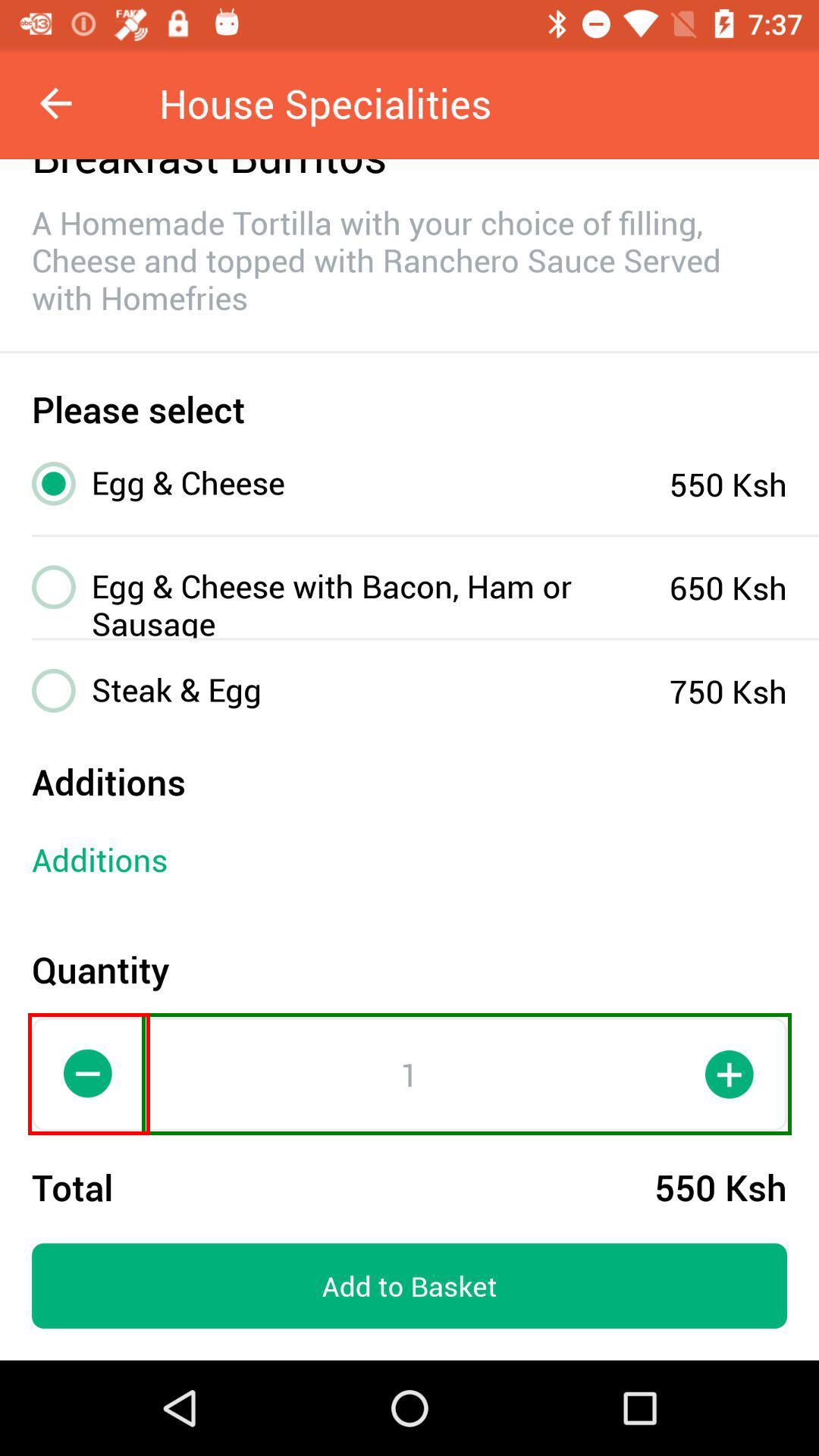}
    }
    \caption{Examples where \toolName{} infers a sub-optimal icon alt-text;
    ref~=~up to 3 WC20 ground-truths. \ref{fig:confusing}'s ground truth is: ``delete item'', 
    ``this button is here if you want more than one item select quantity''.}
    \label{fig:failed_cases}
\end{figure*}

To understand more deeply where \toolName{} fails to generate appropriate icon alt-texts, we randomly chose and manually review 50~examples where the \toolName{}-generated alt-text did not match reference annotations. Following are the four major groups we identified.

\subsubsection{Small Icon Differences Can Trigger Misleading Icon-only Labels}

In some cases an icon looks very similar to other common icons, which misleads \toolName{}'s icon-only label generation. For example, the Figure~\ref{fig:class_effect} down-vote icon (\includegraphics[height=.3cm]{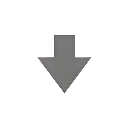}) looks very similar to a common download icon (\includegraphics[height=.3cm]{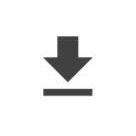}). Gemini's resulting icon-only label in this case confused \toolName{}'s context-based inference enough that it could not recover to a more meaningful alt-text.

\subsubsection{DOM-derived Icon Context May Contain Misleading Terms}

In some cases the DOM-based icon context is very noisy. For example, developers may use naming strategies involving many generic or misleading terms. In the Figure~\ref{fig:context_effect} example, four UI elements' resource-ids contain the term ``hint'' (\code{imgHintBuy}, \code{imgHint50}, \code{imgHintAndroid}, and \code{imgHintFreeze}), which triggered \toolName{} to also include ``hint'' in its alt-text.

\subsubsection{Non-icons} 

In some cases our identification heuristics classified other UI elements as icons (as they are a normally shaped ImageView or ImageButton). For example, the Figure~\ref{fig:not_icon} screen shows a grid of clickable thumbnail images. Since in this work we focus on icons, \toolName{} often produces an unhelpful alt-text in such cases.

\subsubsection{Dubious Ground Truth} 

Some ground truths seem dubious. For example, the second Figure~\ref{fig:confusing} ground truth (``this button is here if you want more than one item select quantity'') should maybe be split, separating ``select quantity'' from the rest. Even after such a hypothetical split the resulting alt-texts appear imprecise or wrong. In any case, for this screen the \toolName{}-inferred alt-text appears more appropriate than the ground truth (while receiving low scores on automated metrics).

\section{Related Work}

\subsection{Necessity and State of UI Accessibility in Mobile Applications}

Accessibility is defined as ``[t]he quality of being easily reached, entered, or used by people with disability''~\cite{iwarsson2003accessibility}. Accessibility in mobile applications is fundamental to ensuring that these technologies are inclusive and can be effectively used by individuals with disabilities, including those with visual, auditory, motor, or cognitive impairments. For instance, visually impaired individuals may rely on screen readers~\cite{talkback, voiceover} to navigate mobile interfaces audibly, while those with motor impairments might require alternative input methods such as voice commands or gestures. Despite significant advancements in technology, mobile applications often fall short of being fully accessible. 

Research has consistently highlighted a range of accessibility barriers that prevent these applications from being fully accessible~\cite{alshayban2020accessibility,vendome2019can, ross2017epidemiology}. An epidemiology-inspired framework by Ross et al. identified multiple determinants that render apps inaccessible, suggesting that even well-designed apps could inadvertently include barriers that limit user engagement~\cite{ross2017epidemiology}. Alshayban et al. found a wide range of accessibility issues in Android apps across 33 different categories, including missing text labels, touch target size, and low contrast~\cite{alshayban2020accessibility}. These studies highlight the persistent challenges in mobile app accessibility and the need for developers to be more aware of accessibility guidelines and best practices. Studies have been conducted to understand developers' perspectives on the issue and how they deal with accessibility in practice~\cite{di2022making}. Their survey reflected that while the developers consider the accessibility guidelines easy to implement, they are still reluctant to implement them. This reluctance suggests that the problem is connected to either a limited understanding of the significance of these guidelines for users with disabilities or a lack of willingness to implement them.

Several international standards and guidelines, such as those from the W3C's Web Content Accessibility Guidelines (WCAG), provide a framework for creating more accessible digital content~\cite{w3c,wcag}. Various companies have also developed their developer guidelines based on established standards, including the Android Accessibility Developer Guidelines~\cite{android_guideline}, Apple’s Accessibility Developer Guidelines~\cite{apple_guideline}, and the IBM accessibility checklist~\cite{ibm_guideline}. These standards offer a range of recommendations aimed at improving support for individuals with diverse disabilities. Developers and designers are encouraged to adhere to these guidelines to enhance the accessibility of their applications. Tools like Lint~\cite{lint} for static analysis within development environments and Accessibility Scanner~\cite{accessibility_scanner} for dynamic analysis on devices offer broader detection capabilities for accessibility issues. These tools, alongside developer training and awareness, play a vital role in bridging the gap between current practices and the ideal state of fully accessible mobile applications.

\subsection{LLMs and VLMs in UI Understanding}

The integration of Large Language Models (LLMs) into UI tasks has transformed how users interact with mobile and web applications, with conversational interfaces becoming increasingly common. Wang et al. demonstrated the potential of LLMs by transforming the mobile UI’s view hierarchy into structured HTML, enabling the models to generate detailed screen summaries and respond to user queries about on-screen elements, thereby simplifying navigation for all users, including those with disabilities~\cite{wang2023enabling}. This marked a foundational step in using LLMs to enhance mobile app usability and accessibility. Building on this, AutoDroid by Wen et al. showcased the advanced capabilities of LLMs in mobile task automation by employing the commonsense knowledge inherent in LLMs along with domain-specific insights from automated app analysis~\cite{wen2024AutoDroid}.

LLMs have also extended their utility to UI and accessibility testing, a critical area for ensuring applications are accessible to users with disabilities. AXNav, introduced by Taeb et al., combines LLMs with pixel-based UI analysis to automate accessibility testing. This system enhances the efficiency of manual testing methods by executing tests based on natural language instructions, thereby broadening the testing scope and scalability~\cite{taeb2024axnav}. Othman et al. have explored how LLMs, particularly ChatGPT, can augment traditional web accessibility evaluations, ensuring thorough and accurate testing procedures~\cite{othman2023fostering}. Similarly, GPTDroid by Liu et al. utilizes GPT-3 to generate dynamic testing scripts that significantly improve bug detection and activity coverage. By treating GUI testing as a question-and-answer task, GPTDroid offers a novel approach to identifying and resolving UI issues~\cite{liu2023chatting}. Additionally, Liu et al.'s QTypist leverages LLMs for automated text input generation in mobile GUI testing, increasing testing accuracy and coverage across diverse applications by integrating with existing GUI testing tools~\cite{liu2023fill}.

Moreover, the recent advancements in Vision-Language Models (VLMs) introduce a new dimension to UI understanding by combining visual and textual information. To address traditional VLMs' underperformance on UI-specific tasks due to a lack of training data tailored to the UI domain, Jiang et al. generated a large dataset of UI screenshots paired with text descriptions and questions using a combination of UI element detection and a Large Language Model (LLM). ILuvUI, created by fine-tuning LLAVA~\cite{llava} with this dataset showcases enhanced capabilities in understanding and interacting with UIs through natural language. Similarly, AMEX dataset by Chai et al. focuses on mobile GUI-control agents and employs GPT-4o to generate detailed descriptions of GUI screens and element functionality~\cite{AMEX_2024}. Both ILuvUI and AMEX employ LLMs in a manner similar to our task of alt-text generation, emphasizing the potential of these models to enrich the dataset creation process and ultimately enhance the accessibility and usability of mobile applications. 

These works demonstrate the growing capability of LLMs and VLMs to understand and interact with complex UIs, proving their potential to significantly improve mobile app usability, accessibility, and testing workflows.

\subsection{Enhancing Mobile UI Accessibility for Users with Visual Impairments}

For individuals with visual impairments, informative alt-text labels are crucial as they offer auditory descriptions of image-based UI elements, enhancing their ability to effectively navigate apps. Despite their importance, the provision of meaningful alt-texts is often neglected by developers, leading to significant accessibility barriers~\cite{alshayban2020accessibility}. Researches underscore the widespread issue of missing and uninformative labels in Android apps, emphasizing the urgent need for automated, informative description generation for UI elements\cite{fok2022large,ross2018examining}. 

Several pioneering studies have aimed to address this by focusing on labeling image-based UI elements. Icons, which are used in most image-based UI elements to convey information and functionality, have been a primary focus. Chen et al. analyzed a large dataset of icons extracted from iPhone apps, highlighting the imbalance between common and long-tail icon types~\cite{chen2022towards}. They further proposed a comprehensive labeling method that combines image classification and few-shot learning, further enhanced by incorporating contextual information from nearby text and modifiers. Their approach primarily categorizes icons based on their class and any detected modifiers, focusing on structural classification. Similarly, Zang et al. introduced a deep learning-based multimodal approach that combines pixel and view hierarchy features to improve icon annotation in mobile applications.~\cite{zang2021multimodal}. Their model significantly enhances icon detection and classification by incorporating both pixel features and textual attributes from view hierarchies. In contrast, our work extends beyond simple classification to generate descriptive alt-text for each icon, providing richer, contextual information that enhances usability for visually impaired users. The most relevant DL approaches of automated alt-text generation are LabelDroid~\cite{chen2020unblind} and Coala~\cite{mehralian2021data}. While LabelDroid and Coala focused on icon labeling through image captioning techniques and incorporated view hierarchy data, their reliance on balanced datasets limited their performance. Our approach overcomes these limitations by functioning effectively with a very limited dataset used for fine-tuning, making it more adaptable to future transitions in UI design.

Recent advancements in Vision-Language Models (VLMs) have further expanded the potential for accessibility in mobile UIs. Spotlight advances mobile UI understanding by utilizing VLMs in a vision-only approach to address tasks such as widget captioning and screen summarization, bypassing the often incomplete or inaccurate view hierarchies~\cite{Spotlight2023}. Chen et al. introduced PaLI, a multimodal model capable of handling both visual and textual inputs, achieving state-of-the-art performance across a variety of tasks such as image captioning, visual question answering (VQA), and image-text retrieval~\cite{chen2022pali}. PaLI is trained on large-scale image-text pairs across multiple languages and performs exceptionally well on a variety of tasks such as image captioning, visual question answering (VQA), and image-text retrieval. The PaLI family’s latest model, PaLI-3~\cite{PaLI-3}, has shown outstanding results in widget captioning, surpassing even its larger predecessor, PaLI-X~\cite{PaLI-X}. ScreenAI~\cite{baechler2024screenai} takes inspiration from these models and combines them with Pix2Struct’s ~\cite{Pix2Struct2023} mechanisms to excel in UI-related tasks like widget captioning and screen summarization. In comparison to these approaches, our model specifically addresses the challenge of partial UI screens, making it more flexible and adaptable to incomplete UI information, which is especially valuable during early development stages when the entire UI may not be fully available.

We evaluate our work against the relevant available models Pix2Struct~\cite{Pix2Struct2023} and PaliGemma~\cite{PaliGemma}.

\section{Conclusions}

User interface icons are essential for navigation and interaction and often lack meaningful alt-text, creating barriers to effective use. Traditional deep learning methods,  while effective in some contexts, generally require extensive and balanced datasets and often struggle when faced with UI icons diversity. Moreover, current Vision-Language Models (VLMs), while sophisticated, do not perform optimally with partial UI screens. We introduced \toolName{}, a novel approach that efficiently generates informative alt-text from partial UI screens, using a relatively small dataset of about 1.4k icons. Our results demonstrated significant improvements in generating relevant alt-text using partial UI information. Integrating \toolName{} into widely used development tools could further enhance its utility by making alt-text generation a standard part of the development workflow, thereby promoting broader adoption and impact.

\section{Data Availability}

All code and data are available on both GitHub~\cite{our_code} and Figshare\footnote{\url{https://figshare.com/s/610e319cb0d4ed663332}}.

\bibliographystyle{ACM-Reference-Format}
\bibliography{ref}

\end{document}